\documentstyle[prl,aps,epsf,multicol,tabularx]{revtex}

\begin{document}
\draft
\title{Creation of solitons and vortices by Bragg reflection of Bose-Einstein
condensates in an optical lattice}
\author{R.G. Scott$^{1}$, A.M. Martin$^{1}$, T.M. Fromhold$^{1}$, S. Bujkiewicz$^{1}$%
, F.W. Sheard$^{1}$, and M. Leadbeater$^{2}$}
\address{$^{1}$School of Physics and Astronomy, University of Nottingham, NG7 2RD, UK}
\address{$^{2}$Department of Physics, University of Durham, South Road, Durham, DH1
3LE, UK}
\date{\today}
\maketitle

\begin{abstract}
We study the dynamics of Bose-Einstein condensates in an optical lattice and harmonic trap. The condensates are set in motion by displacing the trap and initially follow simple semiclassical paths, shaped by the lowest energy band. Above a critical displacement, the condensate undergoes Bragg reflection. For high atom densities, the first Bragg reflection generates a train of solitons and vortices, which destabilize the condensate and trigger explosive expansion. At lower densities, soliton and vortex formation requires multiple Bragg reflections, and damps the center-of-mass motion.
\end{abstract}

\begin{multicols}{2}
Optical lattices (OLs) provide unprecedented control of transport
through the energy bands of periodic quantum systems. This has led
to beautiful experimental demonstrations of Bloch
oscillations\cite{BenD} and quantized Wannier-Stark
ladders\cite{Wilkinson} for non-interacting ultra-cold alkali
atoms. There is also great interest in understanding the behavior
of Bose-Einstein condensates (BECs), formed from interacting
alkali atoms, in
OLs\cite{Jaksch,Berg,Choi,Marzlin,Anders,Holthaus,Josephson,Bronski,Tosi,Morsch,Duttonshock,Greiner,Burger2}.
Predictions\cite{Berg,Choi} that accelerated condensates will
perform Bloch oscillations, whose turning points at the top of the
energy band correspond to successive Bragg reflections,\ have been
confirmed in experiments\cite{Anders,Morsch} on $^{87}$Rb BECs
with equilibrium peak densities $n_{0}\lesssim 10^{14}$ cm$^{-3}$. For $n_{0}
\gtrsim 10^{14}$ cm$^{-3}$, more complex motion has been
observed\cite{Tosi,Morsch}, which cannot be explained by Bragg
reflection or analogous semiclassical models of energy band
transport. Previous numerical studies of condensate dynamics in
OLs have used the one-dimensional (1D) Gross-Pitaevskii
equation\cite{Choi,Holthaus,Tosi}.  They provide invaluable
insights for understanding the mean center-of-mass motion of the
condensate, but have not related this motion to changes in the
{\it internal} structure of the condensate, in particular
dynamical excitations such as solitons and vortices. Producing
such excitations in a controlled way requires state-of-the-art
experimental techniques, which involve manipulating the condensate
phase and/or density
profile\cite{Burger2,Matthews,Burger3,Denschlag,Andvortex}, rotating the
confining trap\cite{Madison,Abo}, moving a laser beam through the
atom cloud \cite{Duttonshock,Raman,Jackson}, or tuning the
inter-atomic interactions\cite{Strecker}.

In this Letter, we show that Bragg reflection provides a new
mechanism for generating solitons and vortices in BECs. Moreover,
these excitations can have a dramatic effect on the evolution of
the atom cloud. At the first Bragg reflection, the condensate
wavefunction is a standing wave with nodes at each maximum in the
OL potential. At each node, the condensate phase changes abruptly
by $\pi$. Bragg reflection therefore imprints atom density and
phase profiles similar to those used to generate solitons in
experiment\cite{Burger2,Matthews,Burger3,Denschlag,Andvortex}. The
effect of this imprinting on the condensate dynamics depends
critically on the atom density.  For condensates with $n_{0}
\gtrsim 10^{14}$ cm$^{-3}$, realized in recent
experiments\cite{Tosi,Morsch}, it leads to the self-assembly of a
chain of stationary solitons, which decay rapidly into vortex
anti-vortex pairs. Strong interactions between the vortices
destabilize the atom cloud, causing it to explode and fragment.
For $n_{0}\lesssim 10^{14}$ cm$^{-3}$, the standing wave formed at
the first Bragg reflection produces no dynamical excitations. But
subsequent Bragg reflections do generate solitons and vortices,
which damp the center-of-mass motion. The dissipation and
instability processes that we identify could play a key role in
the complex dynamics recently observed for high-density
condensates in OLs\cite{Tosi,Morsch}.

%In this Letter, we show that soliton formation is an intrinsic
%feature of condensate dynamics in an OL, because it provides the
%mechanism by which Bragg reflection and Bloch oscillations
%actually occur.  In particular, when the condensate is
%accelerated, the first Bragg reflection proceeds by the
%self-assembly of a chain of stationary solitons, which extends
%across the entire condensate.  The effect of this soliton chain on
%the condensate's subsequent center-of-mass motion depends
%critically on the atom density. For condensates with
%$n_{0}\lesssim 10^{14}$ cm$^{-3}$, the chain disappears rapidly
%after the first Bragg reflection and thereafter does not affect
%the dynamics. But subsequent Bragg reflections produce
%longer-lasting solitons, which decay into vortex anti-vortex
%pairs, thereby damping the center-of-mass motion. For condensates
%with $n_{0} \gtrsim 10^{14}$ cm$^{-3}$,  realized in recent
%experiments\cite{Tosi,Morsch}, the soliton chain has a
%catastrophic effect on the condensate because the associated
%density fluctuations destabilize the atom cloud, causing it to
%explode and fragment.  This new source of instability provides a
%possible explanation for the complex dynamics recently observed
%for high-density condensates in OLs\cite{Tosi,Morsch}.
We consider condensates formed from $N_{A}$ $^{87}$Rb atoms in a
1D OL\ and a three-dimensional harmonic trap. Figure 1(a) shows
the potential energy profile of the OL, $V_{OL}(x)=$ $V_{0}\sin
^{2}(\pi x/d)$, whose depth $V_{0}=$ 23 peV and period $d=$ 397.5
nm are taken from experiment\cite {Tosi}.  The trap frequency for
confinement along the $z-$direction is high enough for the BEC
dynamics to reduce to two-dimensional (2D) motion with potential
energy $V(x,y)=V_{OL}(x)+\frac{1}{2} m(\omega _{x}^{2}x^{2}+\omega
_{y}^{2}y^{2})$, where $m$ is the mass of a single atom and
$\omega _{x}$, $\omega _{y}$ are frequencies of the harmonic
trap. For most of the results presented here,
$\omega _{x}=2\pi \times 50$ rad s$^{-1}$, $\omega _{y}=2\pi
\times 35$ rad s$^{-1}$, $N_{A}=10^{4}$ , and $n_{0}=0.43\times
10^{14}$ cm$^{-3}$ (System A). We also consider a second set of
parameters, $\omega _{x}=2\pi \times 8.7$ rads$^{-1}$, $\omega
_{y}=2\pi \times 90$ rads$^{-1}$, and $N_{A}=3\times 10^{5\text{
}}$(System B), corresponding to recent experiments \cite{Tosi}. In
this case, $n_{0}$ is sufficiently high ($\sim 1.5\times 10^{14}$
cm$^{-3}$) for the standing wave formed during Bragg reflection to have
a particularly dramatic effect on the BEC. For both sets of
parameters, $\omega _{x}$ is small enough to ensure that the
harmonic potential energy variation across each OL period is much
less than the width $\Gamma =10$ peV of the lowest energy band
(Fig. 1(a)). Consequently, the harmonic trap only weakly perturbs
the band structure\cite{Scott}.

We determine the density profile of the condensate by
\begin{figure}
\narrowtext
%\vspace*{0.5cm}
%\epsfysize=9cm
\epsfxsize=6.0cm \centerline{\epsffile{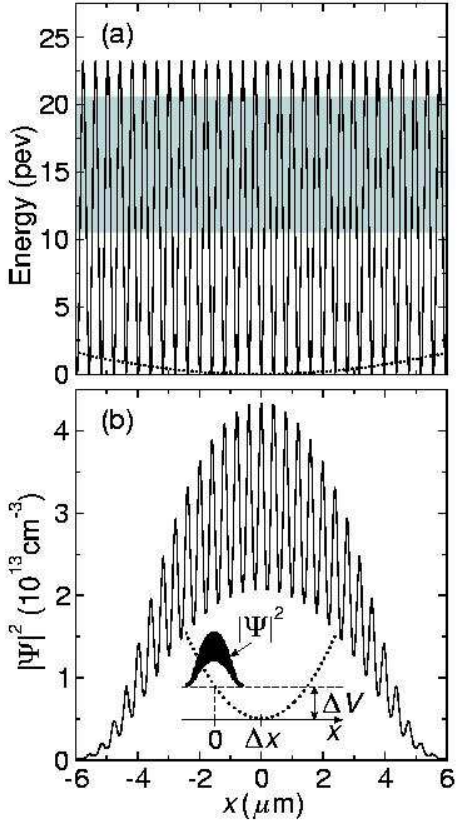}} %6.0cm
\vspace{0.0cm} \caption{(a) Solid curve: potential energy profile %0.15cm
of the OL. Grey rectangle: energy range of the lowest energy band.
Dotted curve: $x-$dependence of the harmonic potential energy. (b)
Initial density profile of System A, along $y=0$. Inset: density
profile and $x-$dependence of harmonic potential energy (dotted)
immediately after trap displacement.}
\end{figure}
\vspace{-0.15cm} %0
\noindent using the Crank-Nicolson method\cite{Choi} to solve the
time-dependent Gross-Pitaevskii equation
\begin{equation}
i\hbar \frac{\partial \Psi }{\partial t} =\left[ \frac{-\hbar
^{2}}{2m}\nabla^2+V(x,y)+\frac{4\pi a\hbar ^{2}}{ m} \left| \Psi
\right|^{2}\right]\Psi,
\end{equation}
where $a=5.4$ nm is the {\it s}-wave scattering length\cite{Burt},
and $\Psi (x,y,t)$ is the condensate wavefunction at time $t$,
normalized so that $\left| \Psi \right|^{2}$ is the number of
atoms per unit volume\cite{Foot1}. The equilibrium  density
profile for the ground state of System A is shown in Fig. 1(b).

At $t=0$, we disrupt the equilibrium of the condensate by suddenly
displacing the harmonic potential through a distance $\Delta x$
along the $x-$axis\cite{Tosi}. Displacing the trap increases the
initial potential energy of the BEC\ by $\Delta V\simeq
\frac{1}{2} m\omega _{x}^{2}(\Delta x)^{2}$ (Fig. 1(b) inset). As
the atoms start to move, this potential energy is converted into
kinetic energy, which determines how far the BEC has accelerated
up the lowest energy band \cite{Scott}. In order for the
condensate to reach the top of the band and therefore undergo
Bragg reflection, $\Delta V$ must be $\geq \Gamma$, which, for
System A, requires $\Delta x\geq (2\Gamma
/m\omega_{x}^{2})^{\frac{1}{2}} =15\mu $m.

We now consider the condensate dynamics obtained from Eq. (1) for
$\Delta x=10\mu $m, below the threshold for
Bragg reflection. Figure 2(a) shows that the mean
\begin{figure}
\narrowtext
%\vspace*{0.5cm}
%\epsfysize=9cm
\epsfxsize=8.5cm \centerline{\epsffile{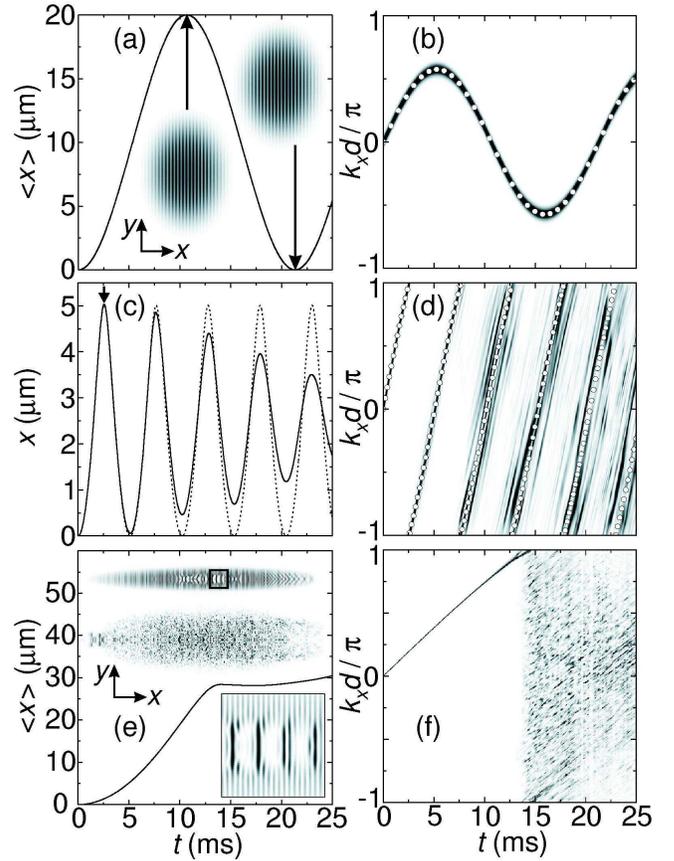}} \vspace{0.1cm} %8.5 0.1
\caption{(a) $\langle x \rangle$ versus $t$ for System A with
$\Delta x=10$ $\mu $m. Insets: grey-scale plots of density (black
high) in $x-y$ plane (axes inset) at $t=$10.7 ms (left) and 21.3
ms (right). (b) Grey-scale plot: $ \left| f(k_{x},t)\right| ^{2}$
(white = 0, black high) for System A with $\Delta x=10\mu $m. Open
circles: points on corresponding semiclassical trajectory
$k_{x}(t)$. (c) Solid curve: $\langle x \rangle$ versus $t$ for
System A with $\Delta x=25\mu $m. Arrow marks first turning point.
Dashed curve: corresponding semiclassical orbit $x(t)$. (d) As
(b), but for $\Delta x=25\mu $m. (e) $\langle x \rangle$ versus
$t$ for System B with $\Delta x=150\mu$m. Insets: grey-scale plots
of density (black high) in $x-y$ plane at $t=$ 13.3 ms (top) and 18
ms (middle). Lower inset shows enlargement of boxed region in upper inset.
(f) As (b), but for System B with $\Delta x=150\mu$m
and omitting the $k_{x}(t)$ curve which, for $ t < 13$ ms, is
indistinguishable from the narrow Fourier distribution.}
\end{figure}
\vspace{-0.8cm}
\noindent (center-of-mass) position of the condensate, $\langle x
\rangle$, undergoes simple periodic motion, bounded by the
harmonic trap. The internal structure of the BEC is unaffected by
this motion, being the same at $t=10.7$ ms (Fig. 2(a) left inset)
and $t=21.3$ ms (Fig. 2(a) right inset) as at $t=0$. To determine
how the condensate moves in reciprocal space, we calculate the
Fourier transform of $\Psi (x,0,t)$. The Fourier power, $\left|
f(k_{x},t)\right| ^{2}$, corresponding to wavenumber $k_x$,
remains narrow and changes periodically as $t$ increases
(grey-scale plot in Fig. 2(b)). Since the condensate's internal
structure  does not change with $t$ when $\Delta x=10 \mu $m, the
form of $\left| f(k_{x},t)\right| ^{2}$ and the corresponding $
\langle x \rangle $ versus $t$ curve (Fig. 2(a)) can be
understood by considering the motion of a single point particle in
\begin{figure}
\narrowtext
%\vspace*{0.5cm}
%\epsfysize=9cm
\epsfxsize=8.5cm \centerline{\epsffile{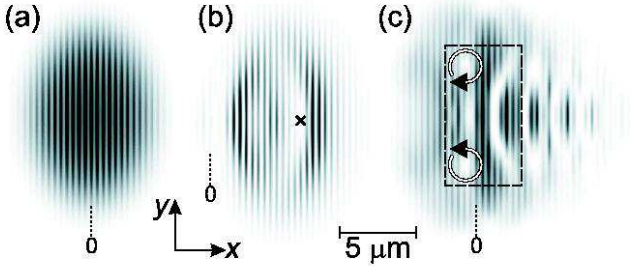}} \vspace{0.0cm} %8.5 0.1
\caption{Grey-scale plots of  density (white = 0, black high) in
$x-y$ plane (axes inset) for System A with $\Delta x=25$ $\mu $m
and $t=0$ ms (a), 7.5 ms (b), 10.7 ms (c). Plots are
symmetrical under reflection about $y=0$. Vertical dotted lines
indicate $x=0$ in each case. Horizontal bar shows scale. Cross in
(b) marks center of a soliton.  Region within dashed box in
(c) is shown in Fig. 5(a).}
\end{figure}
\vspace{-0.2cm} %0
\noindent the lowest
energy band. The single particle trajectories $x(t)$ and
$k_{x}(t)$ in real and reciprocal space are determined by
the semiclassical equations of motion $dx/dt=\hbar
^{-1}dE(k_{x})/dk_{x}$ and $dk_{x}/dt=\hbar^{-1} F_{x}$\cite{A},
where $E(k_{x})$ is the energy-wavenumber dispersion relation for
the band, and $F_{x}=-m\omega _{x}^{2}x$ is the harmonic restoring
force along the $x-$direction. In Fig. 2(b), the Fourier power
(grey-scale plot) is concentrated along the single-particle
$k_{x}(t)$ curve (white circles). The corresponding real-space
trajectory, $x(t)$, is indistinguishable from the plot of $\langle
x \rangle$ versus $t$ shown in Fig. 2(a).

When $\Delta x$ is increased to $25 \mu $m, above the threshold
for Bragg reflection, the mean $x$ position of the condensate,
determined from Eq. (1), performs damped periodic motion (solid
curve in Fig. 2(c)). In Fig. 2(d), the grey-scale plot of $\left|
f(k_{x},t)\right| ^{2}$ shows that the condensate's mean $k_{x}$
value increases approximately linearly with $t$ and reaches the
Brillouin zone boundary at 2.6 ms. At this time, the condensate
undergoes Bragg reflection, which produces the first (arrowed)
turning point in Fig. 2(c). The quantum calculations of $\langle x
\rangle $ and $\left| f(k_{x},t)\right|^{2}$ deviate rapidly from
the corresponding semiclassical trajectories, $x(t)$ and
$k_{x}(t)$, shown respectively by the dashed curve and open
circles in Figs. 2(c) and (d). For $t\gtrsim 7.5$ ms, the
oscillations in $\langle x \rangle$ are damped and multiple peaks
appear in $\left| f(k_{x},t)\right| ^{2}$, which spreads through
the Brillouin zone. This deviation from single-particle behavior
indicates that the BEC's center-of-mass motion is strongly
affected by changes in its internal structure. Key stages in the
evolution of the 2D density profile are shown in
 Fig. 3. As $t$ increases from $0$ (Fig. 3(a)), the
density minima deepen and fall to zero at the first Bragg
reflection, which we now analyze in detail.

The lower curves in Figs. 4(a), (b), and (c) show $\left| \Psi
(x,0,t)\right| ^{2}$ just before $(t=2$ ms$)$, at $(t=2.6$ ms$)$,
and just after $(t=3$ ms$)$ the first Bragg reflection. The upper
curves show the wavefunction phase, $\phi (x)$, modulo 2$\pi $.
Just before reflection (Fig. 4(a)), the density near
the center of the BEC has a minimum value of
$\sim 10^{13}$ cm$^{-3}$, which is
approximately half that at $t=0$ (Fig.
\begin{figure}
\narrowtext
%\vspace*{0.5cm}
%\epsfysize=9cm
\epsfxsize=8.5cm \centerline{\epsffile{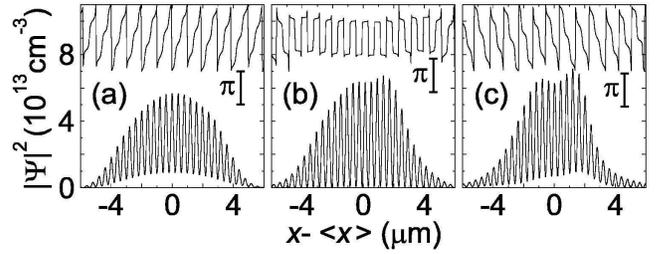}} \vspace{0.05cm} %8.5 0.075
\caption{Lower curves:  density profiles along $y=0$ for the
condensate in System A with $\Delta x=25$ $\mu $m and $t=2$ ms
(a), 2.6 ms (b), 3 ms (c).
Upper curves show $\phi(x)$ modulo 2$ \pi $, with vertical scale
indicated by bars of length $\pi $.}
\end{figure}
\vspace{-0.45cm} \noindent 1(b)). The local velocity %-0.25
along the $x-$direction, $v_{x}=(\hbar /m)d\phi /dx$, is $>0$
throughout the condensate. At the point of Bragg reflection (Fig.
4(b)), the density minima fall to zero at each peak in
$V_{OL}(x)$. At each zero, $\phi $ changes
abruptly by $\pi $ (upper curve in Fig. 4(b)). Away from the
discontinuities, $d\phi /dx\simeq 0$, indicating that the BEC is
at rest. This variation of density and phase demonstrates that a
standing wave forms at the point of Bragg reflection.
In recent experiments, laser illumination was used to produce similar individual
density minima and/or $\pi$ phase
shifts\cite{Burger2,Matthews,Burger3,Denschlag,Andvortex}, which subsequently
evolved into dark solitons. By analogy, the standing wave
%formed at Bragg reflection
might be expected to generate a chain of stationary solitons, each of width
$w \approx \left( \pi a n_{M} \right)^{- \frac{1}{2}} $, where $n_{M}$ is the
local mean atom density\cite{Burger2}. At the first Bragg reflection,
$n_{M} \simeq 3\times 10^{13}$ cm$^{-3}$ near the center of the condensate in
System A (see Fig. 4(b)), and so $w \simeq 3.5d$\cite{Tosi}. Since $w$ is much
larger than the width ($\sim d$) of the density minima in the standing wave,
the first Bragg reflection does not produce solitons in System A. Instead, after
reflection, the density minima rise away from zero and $d\phi /dx$ becomes
negative for all $x$, as the condensate starts to move from right to left.

We now consider the condensate motion for $t\gtrsim 5$ ms.
Figure 3(b) shows the density profile at the second
Bragg reflection, when $t=7.5$ ms. Again a standing wave is formed,
which creates nodal lines in the density profile (white stripes in Fig. 3(b))
at each maximum in $V_{OL}(x)$.
However, in contrast to the first Bragg reflection, the standing wave now disrupts the internal structure of the condensate sufficiently to allow a soliton,
marked by the cross in Fig. 3(b), to form across several OL periods.
Subsequent Bragg reflections generate more
solitons, which have a pronounced effect on the condensate's
internal structure and center-of-mass motion.
To illustrate this,
Fig. 3(c) shows the density profile at $t=10.7$ ms. For $x>0$,
there are three extended solitons (white crescent shapes), whose
wavefronts have been curved by refraction originating from the
non-uniform density\cite{Duttonshock}. This refraction is the
precursor of snake instability\cite{Duttonshock}, which causes the
solitons to decay into two vortices with opposite circulation,
like those enclosed by arrows in Fig. 3(c). At the center of each
vortex, $\left| \Psi \right| ^{2}$
\begin{figure}
\narrowtext
%\vspace*{0.5cm}
%\epsfysize=9cm
\epsfxsize=8.5cm \centerline{\epsffile{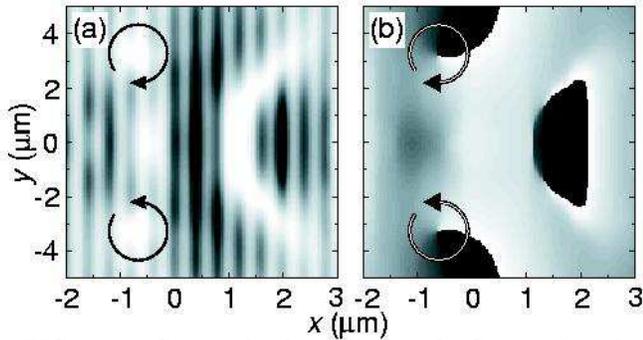}} \vspace{-0.05cm} %8cm 0.3cm
\caption{(a) Grey-scale plot of  density for System A within
dashed box in Fig. 3(c) (white = 0, black high). Arrows show
direction of circulation around vortices. (b) Grey-scale plot of
$\phi $ (white = 0, black = 2$ \pi $). $\Delta x=25$ $\mu $m and
$t=$ 10.7 ms.}
\end{figure}
\vspace{-0.15cm} %0
\noindent $=0$. The vortices can be seen
more clearly in Figs. 5(a) and (b), which show respectively
enlargements of the density profile within the dashed box in Fig.
3(c), and the corresponding phase. The soliton represented by the white
crescent in Fig. 5(a) appears as a dark island towards the
right of Fig. 5(b). At the left-hand edge of this island, the
phase changes abruptly from $\pi /4$ (light grey) to $3\pi /4$
(dark grey). Around the two vortices, the phase (Fig. 5(b))
changes continuously from 0 (white)\ to 2$\pi $ (black),
indicating quantized circulation in the direction of the arrows.
Vortex formation is the main cause of damping in the
center-of-mass motion (Fig. 2(c)). A crucial aspect of this
damping mechanism is that the soliton formation and subsequent
vortex shedding occur when the condensate is almost {\it at rest}.
It is therefore {\it fundamentally different} to the phonon
emission process used to interpret experiments on System
B\cite{Tosi}, which occurs when $v_x$ {\it exceeds} a critical
value of $\sim 5$ mm s$^{-1}$.
It is also unrelated to the damping found in 1D simulations of Bloch-oscillating
condensates\cite{Trombett}, which cannot include the effects of vortex formation.

We now relate our calculations to the experiments on System
B\cite{Tosi}. Figures 2(e) and (f) show the time evolution of $
\langle x \rangle$ and $\left| f(k_{x},t)\right| ^{2}$ for this
system, after a large trap displacement of $150\mu $m. As in
System A, the first Bragg reflection generates a density node and
associated $\pi$ phase shift at each maximum in $V_{OL}(x)$. But
since $n_{M}$ is much larger in System B (  $\sim3.3\times10^{14}$
cm$^{-3}$ near the BEC center at the first Bragg reflection), $w
\simeq d$. Since $w$ is so closely matched to the width of the
density minima in the standing wave, Bragg reflection causes the
self-assembly of $\sim 10$ stationary solitons, which form a chain
across the central third of the condensate. Figure 2 (e) shows the
compact cigar-shaped density profile of the condensate just after
the first Bragg reflection (upper inset).  The region within the
box is shown enlarged in the lower inset of Fig. 2(e), which
reveals three stationary solitons (extended white areas). The time
taken for solitons to develop following phase imprinting is
proportional to the distance $l$ over which the phase changes by
$\pi$\cite{Burger2}. In a standing wave, $l \simeq 0$, and so the
first Bragg reflection and the formation of the soliton chain
occur almost instantaneously.
%less than 0.5 ms after the first Bragg reflection. % CHANGE HERE
%Each soliton decays rapidly into vortex anti-vortex pairs.  There are strong
%repulsive forces between neighboring solitons, because they have a
%phase difference of %$\pi$\cite{Strecker,Gordon}, and also between vortices of
%opposite circulation.
The solitons decay rapidly into two chains of vortex anti-vortex
pairs, which form a complex interacting system. The interactions
create a massive internal strain, which causes the BEC to explode
laterally (perpendicular to the $x$-axis), resulting in the
diffuse and fragmented atom density profile shown in the middle
inset of Fig. 2(e) \cite{New}. The explosion has a dramatic effect
on the $k_{x}-$distribution of the atoms (Fig. 2(f)), which is
initially extremely narrow but, at the point of Bragg reflection,
spreads through the whole Brillouin zone. This could account for
the broad momentum distribution observed when high-density
condensates undergo {B}ragg reflection\cite{Morsch}.

In summary, we have investigated how Bragg reflection affects the
internal structure and center-of-mass-motion of condensates
accelerating through an OL. When the atom density is high enough
to ensure that $w \lesssim d$, the density zeros and $\pi$ phase
shifts imprinted by the first Bragg reflection generate a train of
stationary solitons, which decay rapidly into vortex anti-vortex
pairs.  Strong interactions between the ensemble of vortices have
a catastrophic effect on the condensate, causing it to undergo
explosive expansion. This dynamical regime is a unique feature of
condensates in an OL and should be experimentally accessible in
existing systems\cite{Tosi,Morsch}. For lower atom densities,
soliton formation requires multiple Bragg reflections. The
subsequent decay of the solitons into vortex anti-vortex pairs
provides a new dissipation mechanism, which could contribute to
the damping of the center-of-mass oscillations observed in
experiment\cite{Tosi}. 
\end{multicols}

\vspace{-0.6cm}
\begin{references}
\vspace{-1.7cm}
\bibitem{BenD}  M. Ben Dahan et al, Phys. Rev. Lett. {\bf 76}, 4508 (1996).

\bibitem{Wilkinson}  S.R. Wilkinson et al, Phys. Rev. Lett. {\bf 76}, 4512
(1996).

\bibitem{Jaksch}  D. Jaksch et al, Phys. Rev. Lett. {\bf 81}, 3108 (1998).

\bibitem{Berg}  K. Berg-S\o rensen and K. M\o lmer, Phys. Rev. A {\bf 58},
1480 (1998).

\bibitem{Choi}  D.I. Choi and Q. Niu, Phys. Rev. Lett. {\bf 82}, 2022
(1999).

\bibitem{Marzlin}  K.P. Marzlin and W. Zhang, Phys. Rev. A {\bf 59},
2982 (1999).

\bibitem{Anders}  B.P. Anderson and M.A. Kasevich, Science {\bf 282}, 1686
(1998).

\bibitem{Holthaus}  M. Holthaus, J. Opt. B: Quantum Semiclass. Opt. {\bf 2},
589 (2000).

\bibitem{Josephson}  F.S. Cataliotti et al, Science {\bf 293}, 843 (2001).

\bibitem{Bronski}  J.C. Bronski et al,
Phys. Rev. Lett. {\bf 86}, 1402 (2001).

\bibitem{Tosi}  S. Burger et al, Phys. Rev. Lett. {\bf 86}, 4447 (2001).

\bibitem{Morsch}  O. Morsch et al, Phys. Rev. Lett. {\bf 87}, 140402 (2001).

\bibitem{Duttonshock}  Z. Dutton et al, Science {\bf 293}, 663 (2001).

\bibitem{Greiner}  M. Greiner et al, Phys. Rev. Lett. {\bf 87}, 160405
(2001); Nature {\bf 415}, 39 (2002).

\bibitem{Burger2}  S. Burger et al, Phys. Rev. A. {\bf 65}, 043611 (2002).

\bibitem{Matthews}  M.R. Matthews et al, Phys. Rev. Lett. {\bf 83}, 2498
(1999).

\bibitem{Burger3}  S. Burger et al, Phys. Rev. Lett. {\bf 83}, 5198 (1999).

\bibitem{Denschlag}  J. Denschlag et al, Science {\bf 287}, 97 (2000).

\bibitem{Andvortex}  B.P. Anderson et al, Phys. Rev. Lett. {\bf 86}, 2926
(2001).

\bibitem{Madison}  K.W. Madison et al, Phys. Rev. Lett. {\bf 84}, 806 (2000).

\bibitem{Abo}  J.R. Abo-Shaeer et al, Science {\bf 292}, 476 (2001).

\bibitem{Raman}  C. Raman et al, Phys. Rev. Lett. {\bf 83}, 2502 (1999).

\bibitem{Jackson}  B. Jackson et al, Phys. Rev. A {\bf 61}, 051603(R) (2000).

\bibitem{Strecker} K.E. Strecker et al, Nature {\bf 417}, 150
(2002).

\bibitem{Scott}  R.G. Scott et al, Phys. Rev. A, to be published.

\bibitem{Burt}  E.A. Burt et al, Phys. Rev. Lett. {\bf 79}, 337 (1997).

\bibitem{Foot1}  For the densities considered here, inter-atomic interactions
have a negligible effect on the OL potential and energy band
structure.

\bibitem{A} See, for example, {\it Solid State
Physics}, by N.W. Ashcroft and N.D. Mermin (Holt, Rinehart, and
Winston, 1976).

\bibitem{Trombett} A. Trombettoni and A. Smerzi,
Phys. Rev. Lett. {\bf 86}, 2353 (2001).

\bibitem{New} This trigger of explosion might also play
a role in the dynamical instability proposed in B. Wu and Q. Niu,
Phys. Rev. A {\bf 64}, 061603 (2001).

\newpage
\end{references}
\end{document}